\documentclass[conference]{IEEEtran}
\IEEEoverridecommandlockouts
\usepackage{cite}
\usepackage{amsmath,amssymb,amsfonts}
\usepackage{algorithmic}
\usepackage{graphicx}
\newtheorem{proposition}{\underline{Proposition}}
\newtheorem{lemma}{\underline{Lemma}}
\usepackage{textcomp}
\usepackage{xcolor}
\usepackage[hidelinks]{hyperref}
\newcommand{\mv}[1]{\mbox{\boldmath{$ #1 $}}}
\def\BibTeX{{\rm B\kern-.05em{\sc i\kern-.025em b}\kern-.08em
		T\kern-.1667em\lower.7ex\hbox{E}\kern-.125emX}}
\begin{document}
	
	\title{Secure Integrated Sensing and Communication 
		Exploiting Target Location Distribution}
	\author{\IEEEauthorblockN{Kaiyue Hou and Shuowen Zhang}
		\IEEEauthorblockA{{Department of Electronic and Information Engineering, The Hong Kong Polytechnic University} \\
			Email: kaiyue.hou@connect.polyu.hk, shuowen.zhang@polyu.edu.hk}}
	\maketitle
	
	\begin{abstract}
		In this paper, we study a secure integrated sensing and communication (ISAC) system where one multi-antenna base station (BS) simultaneously serves a downlink communication user and senses the location of a target that may potentially serve as an eavesdropper via its reflected echo signals. Specifically, the location information of the target is \emph{unknown} and \emph{random}, while its \emph{a priori distribution} is available for exploitation. First, to characterize the sensing performance, we derive the \emph{posterior Cram\'er-Rao bound (PCRB)} which is a lower bound of the mean squared error (MSE) for target sensing exploiting prior distribution. Due to the intractability of the PCRB expression, we further derive a novel approximate upper bound of it which has a closed-form expression. Next, under an artificial noise (AN) based beamforming structure at the BS to alleviate information eavesdropping and enhance the target's reflected signal power for sensing, we formulate a transmit beamforming optimization problem to maximize the worst-case secrecy rate among all possible target (eavesdropper) locations, under a sensing accuracy threshold characterized by an upper bound on the PCRB. Despite the non-convexity of the formulated problem, we propose a two-stage approach to obtain its \emph{optimal solution} by leveraging the semi-definite relaxation (SDR) technique. Numerical results validate the effectiveness of our proposed transmit beamforming design and demonstrate the non-trivial trade-off between secrecy performance and sensing performance in secure ISAC systems.
	\end{abstract}
	
	\section{Introduction}
	The inherent broadcast nature of wireless communication exposes it to various risks such as eavesdropping and jamming, posing a significant threat to communication security in wireless applications. Traditional security techniques based on cryptographic approaches are difficult to distribute in large-scale heterogeneous networks due to the challenge of managing secret keys. On the other hand, physical-layer security solutions that exploit the unique wireless channel characteristics to achieve secure transmission without keys have emerged as promising approaches \cite{bib1}. For example, artificial noise (AN) \cite{bib2} has been proposed as an effective method to mitigate the amount of information leakage to the eavesdropper by adding extra AN beams at the transmitter, which has been widely investigated in recent years \cite{bib3,bib5,bib6}.
	
	\begin{figure}
		\centering
		\includegraphics[scale=0.6]{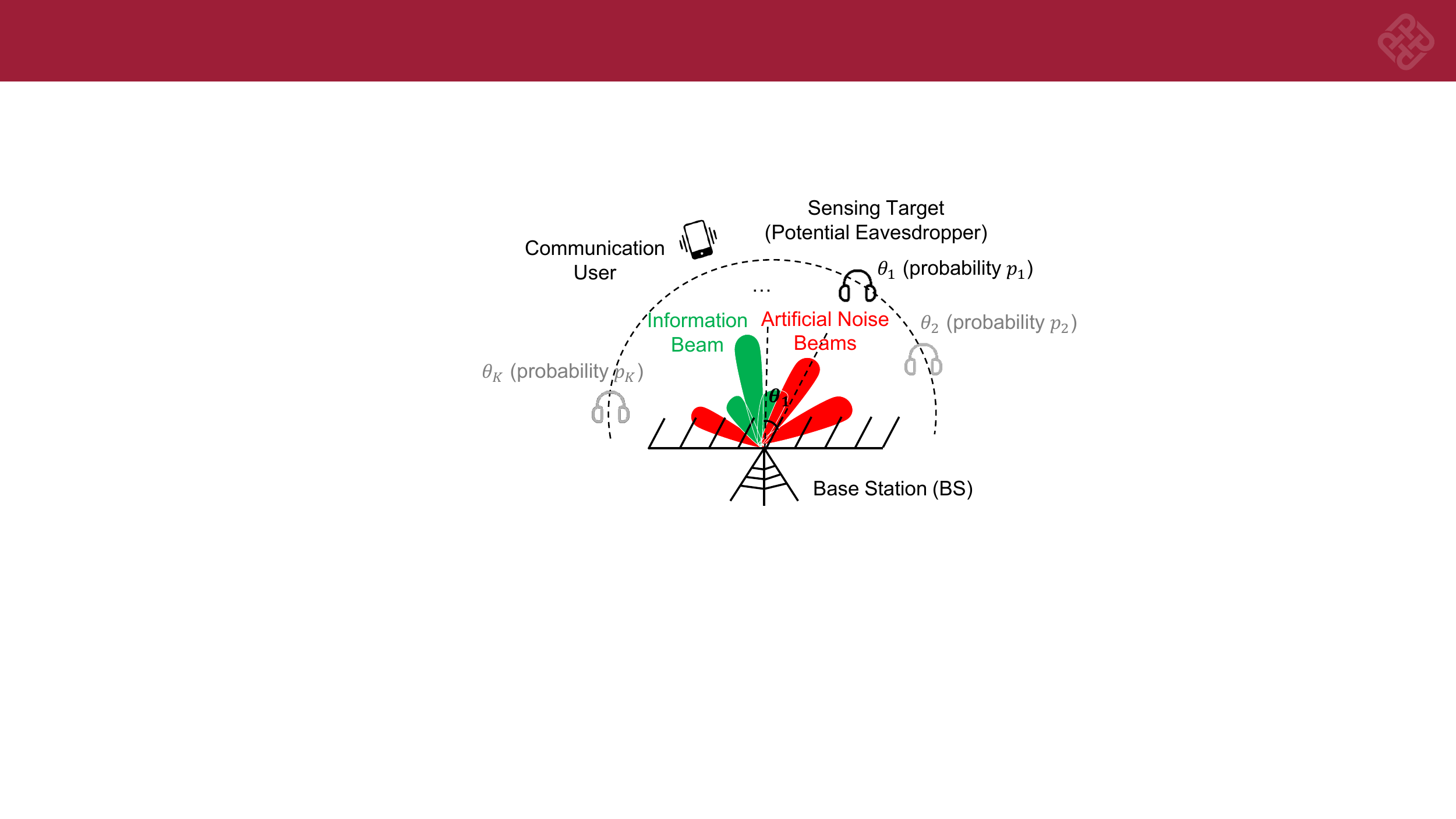}
		\vspace{-3mm}
		\caption{Illustration of a secure ISAC system with random target location.}
		\centering
		\label{Fig1_system}
		\vspace{-8mm}
	\end{figure}
	Along this line, most existing works focused on the scenario where the channel between the base station (BS) and the eavesdropper is known for transmit signal design. However, in practice, the eavesdropping channel or even the location of the eavesdropper may be unknown. To estimate the locations of targets including eavesdroppers, integrated sensing and communication (ISAC) \cite{bib7} has recently emerged as a promising technology, where the transmit signals at the BS can be used to simultaneously serve communication users and perform target sensing. Nevertheless, most existing works on ISAC focused on the target detection and tracking towards a given location, while the case for sensing a target at an unknown location has not been thoroughly investigated. In \cite{bib8}, a novel iterative approach was proposed to successively refine the sensing performance given an uncertain target location. However, the signal design at the BS in each iteration was still focused on improving the sensing performance corresponding to a given location. On the other hand, it is worth noting that the distribution of the target may be known \emph{a priori} via exploring empirical observations or target movement pattern, which can be exploited to enhance the sensing performance. To the best of our knowledge, transmit signal design for ISAC exploiting the distribution of unknown and random target location still remains an open problem, especially for the more challenging case where the target may serve as a potential eavesdropper.
	
	Motivated by the above, we consider a secure ISAC system with a multi-antenna BS, a single-antenna communication user, and a sensing target which may serve as an eavesdropper. The target's exact location is \emph{unknown} and \emph{random}, while its distribution is available to be exploited. First, we introduce a novel \emph{posterior Cram\'er-Rao bound (PCRB)} based method to characterize a lower bound of the mean squared error (MSE) exploiting prior distribution information. Note that in contrast to the Cram\'er-Rao bound (CRB), PCRB does not depend on the exact location of the target. We then derive a novel approximate upper bound of the PCRB in a tractable closed form. Next, considering an AN-based transmit beamforming structure, we formulate the beamforming optimization problem to maximize the worst-case secrecy rate among all possible target (eavesdropper) locations, subject to a requirement on the sensing accuracy characterized by an upper bound on the PCRB. By leveraging the semi-definite relaxation (SDR) technique, we obtain the \emph{optimal solution} to the formulated problem. It is shown via numerical results that our proposed design achieves superior secrecy and sensing performance over various benchmark schemes, due to the smart exploitation of the target (eavesdropper) location distribution.
	
	\section{System Model}
	We consider a secure ISAC system where a BS equipped with $N_t\geq 1$ transmit antennas and $N_r\geq 1$ co-located receive antennas serves a single-antenna communication user in the downlink. Moreover, the BS aims to sense the location of a target which serves as a potential \emph{eavesdropper} via the received echo signals reflected by the target.\footnote{Note that although the target may serve as an eavesdropper and potentially possess active sensing capability, we consider device-free passive sensing since the target will not proactively share its location with the BS.} The exact location information of the target is \emph{unknown} and \emph{random}, while its distribution is available to be exploited as prior information. Specifically, we assume that the target has $K\geq 1$ possible locations, as illustrated in Fig. \ref{Fig1_system}. For ease of revealing fundamental insights, we consider a two-dimensional (2D) coordinate system, where each $k$-th location has the same distance $r$ in meters (m) and a distinct angle $\theta_k\in [-\pi,\pi)$ with respect to the reference point at the BS. The common distance $r$ is assumed to be known \emph{a priori},\footnote{The range information $r$ can be obtained by exploiting empirical observations or estimated \emph{a priori} using e.g., time-of-arrival (ToA) methods.} while the probability for the target to be located at the $k$-th possible angle is denoted by $p_k\in [0,1]$, with $\sum_{k=1}^K p_k=1$. The probability mass function (PMF) of $\theta$ is thus given by
	\begin{align}\label{PMF}
		p_\Theta(\theta)=
		\begin{cases}
			p_k,\ &\mathrm{if}\ \theta=\theta_k,\ k=1,...,K,\\
			0,\ &\mathrm{otherwise}.
		\end{cases}
	\end{align}
	
	We consider a challenging scenario for secrecy communication where the target has a line-of-sight (LoS) channel with the BS, and the downlink eavesdropping channel denoted by $\mv{h}_E^H(\theta)\in \mathbb{C}^{1\times N_t}$ is \emph{unknown} due to the unknown target's angle $\theta$. On the other hand, the channel from the BS to the user denoted by $\mv{h}^H\in \mathbb{C}^{1\times N_t}$ is assumed to be perfectly known at the BS. Furthermore, we assume $\mv{h}_E^H(\theta_k)$'s and $\mv{h}$ are linearly independent, which can hold for various user channel models including the LoS model (with a distinct user angle) and random Rayleigh fading model.
	
	Our objective is to achieve high-quality secrecy communication and sensing performance by optimizing the BS transmit signals via smart exploitation of the \emph{distribution information} about the eavesdropping target's location. Specifically, we aim to maximize the worst-case secrecy rate corresponding to the most favorable eavesdropping location, while ensuring a sensing accuracy of the eavesdropping target's location, which will facilitate more tailored signal designs with further improved secrecy performance in future communication instances. 
	
	To this end, we introduce an \emph{AN-based beamforming design}, where the transmitted signal vector is the superposition of an information beam and $K$ AN beams. Denote $s\!\sim\! \mathcal{CN}(0,1)$ as the information symbol for the user, and ${\mv{w}}\!\in\! \mathbb{C}^{N_t\times 1}$ as the information beamforming vector. We further denote $\mv{v}_k\!\in\! \mathbb{C}^{N_t\times 1}$ as the $k$-th AN beamforming vector, and ${s}_k\!\sim\! \mathcal{CN}(0,1)$ as the $k$-th independent AN signal which is also independent with $s$. The transmitted signal vector is thus given by
	\vspace{-2mm}
	\begin{equation}\label{transmit_signal}
		\mv{x} = \mv{w}s + \sum_{k=1}^K\mv{v}_ks_k,
		\vspace{-2mm}
	\end{equation}
	Let $P$ denote the transmit power constraint, which \hbox{yields} $\mathbb{E}[\|{\mv{x}}\|^2]=\|{\mv{w}}\|^2+\sum_{k=1}^K\|{\mv{v}}_k\|^2\leq P$. Note that the motivation for the AN-based approach is two-fold. Firstly, by introducing additional Gaussian-distributed noise signals which are the worst-case noise for eavesdropping, the received signal-to-interference-plus-noise ratio (SINR) at the potential eavesdropper can be decreased, thus enhancing the \emph{communication secrecy}. Secondly, the extra AN beams provide more design flexibility in strengthening the echo signals from possible target locations, thus enhancing the \emph{sensing accuracy}. 
	
	Based on (\ref{transmit_signal}), the received signal at the user is given by
	\vspace{-2mm}\begin{equation}
		y = \mv{h}^{H}\mv{x}+z={\mv{h}}^H{\mv{w}}s+{\mv{h}}^H\sum_{k=1}^K{\mv{v}}_ks_k+z,\label{con: RsignalUser}	\vspace{-2mm}
	\end{equation}
	where $z\sim\mathcal{C}\mathcal{N}\left( {0,\sigma^{2}}\right)$ denotes the circularly symmetric complex Gaussian (CSCG) noise at the user receiver. The SINR at the user receiver is thus given by
	\vspace{-1mm}\begin{equation}
		{\rm{SINR}} = \frac{| {\mv{h}^{H}\mv{w}} |^{2}}{\sum_{k=1}^K|{\mv{h}^{H}\mv{v}_k} |^{2} + \sigma^{2}}.\label{con: SINR CU}	\vspace{-1mm}
	\end{equation}
	The received signal at the potential eavesdropper is given by
	\vspace{-1mm}\begin{equation}
		\!\!\!	{y_{E}}({\theta})\! =\! {\mv{h}}_E^H(\theta){\mv{x}}\!+\!z_E\!=\!{\mv{h}}_E^H(\theta){\mv{w}}s\!+\!{\mv{h}}_E^H(\theta)\sum_{k=1}^K\mv{v}_ks_k\!+\!z_E,\!\!\! \label{con: RsignalEaves}
		\vspace{-1mm}\end{equation}
	where $z_{E}\!\sim\!\mathcal{C}\mathcal{N}({0,\sigma_{E}^{2}})$ denotes the CSCG noise at the eavesdropper receiver. By noting that the BS-eavesdropper channel $\mv{h}_E^H(\theta)$ follows the LoS model and considering a uniform linear array (ULA) at the BS, we have ${\mv{h}}_E^H(\theta)=\frac{\sqrt{\beta_0}}{r}{\mv{a}}^H(\theta)$, where $\beta_0$ denotes the reference channel power at 1 m; ${\mv{a}}^H(\theta)=[e^{- j\pi\Delta(N_{t} - 1){\sin\theta}}, e^{- j\pi\Delta(N_{t} - 3){\sin\theta}},\dots, e^{j\pi\Delta(N_{t} - 1){\sin\theta}}]$ denotes the steering vector at the BS transmit array, with $\Delta$ denoting the antenna spacing over wavelength ratio. Hence, the SINR at the potential eavesdropper can be expressed as
	\begin{align}
		{\rm{SINR}}_E(\theta)\! =\! \frac{| {\mv{a}^{H}(\theta)\mv{w}} |^{2}}{\sum_{k=1}^K| {\mv{a}^{H}(\theta)}\mv{v}_k |^{2}\! +\! \frac{\sigma_E^2r^2}{\beta_0}},\theta\!\in\! \{\theta_1,...,\theta_K\}.
	\end{align}
	The achievable secrecy rate at the user when there exists an eavesdropper at location $k$ with angle $\theta_k$ is given by \cite{bib9}:
	\begin{align}\label{cons: r_0}
		\!\!\!\!R_k\!=[\log_2(1\!+\!\mathrm{SINR})\!-\!\log_2(1\!+\!\mathrm{SINR}_E(\theta_k))]^+,\forall k\!\!
	\end{align}
	in bps/Hz, where $[a]^+=\max\{a,0\}$. The worst-case achievable secrecy rate among all possible eavesdropper locations is thus given by $R=\underset{k=1,...,K}{\min}\ R_k$.
	
	Besides reaching the user and eavesdropper receivers, the transmit signal will be reflected by the target. Let $\alpha\in \mathbb{C}$ denote the radar cross section (RCS) coefficient, which is generally an \emph{unknown} and \emph{deterministic} parameter. Let ${\mv{b}}(\theta)\!=\![e^{-j\pi\Delta(N_{r}-1)\sin\theta}, e^{-j\pi\Delta(N_{r}-3)\sin\theta},\dots,e^{j\pi\Delta(N_{r}-1)\sin\theta}]^H$ denote the steering vector at the BS receive array. Thus, the received echo signal at the BS receive antennas is given by
	\begin{align} \label{cons: y_R}
		\mv{y}_{R} = \frac{\beta_0}{r^2}\mv{b}(\theta)\alpha\mv{a}^{H}(\theta)\mv{x} + \mv{z}_{R} \triangleq\beta 
		\mv{M}(\theta)\mv{x}+\mv{z}_R,
	\end{align}
	where $\beta\overset{\Delta}{=}\frac{\beta_0}{r^2}\alpha$ denotes the overall reflection coefficient including the two-way channel gain and RCS; 
	${\mv{z}}_R\!\sim\!\mathcal{CN}(\mv{0},\sigma_R^2{\mv{I}}_{N_r})$ denotes the CSCG noise vector at the BS receive antennas; and ${\mv{M}}(\theta)\overset{\Delta}{=}\mv{b}(\theta)\mv{a}^{H}(\theta)$. 
	
	In the next section, we aim to characterize the performance of estimating $\theta$ based on the received signal vector in (\ref{cons: y_R}). 
	
	\section{Sensing Performance Characterization Exploiting Prior Distribution Information}
	Notice from (\ref{cons: y_R}) that the overall reflection coefficient $\beta=\beta_R+j\beta_I$ is also an unknown (and deterministic) parameter, which thus also needs to be estimated to obtain an accurate estimation of $\theta$. Let ${\mv{\omega}}=[\theta,\beta_R,\beta_I]^T$ denote the collection of unknown parameters to be estimated. With the prior distribution information of $\theta$ available for exploitation, we propose to employ \emph{PCRB} as the performance metric, which characterizes a global lower bound of the MSE of unbiased estimators exploiting prior information. To the best of our knowledge, most classic PCRB derivation methods are suitable for estimation parameters with continuous and differentiable probability density functions (PDFs) \cite{bib10}. For consistence, we propose to approximate the discrete PMF in (\ref{PMF}) with a continuous Gaussian mixture PDF given by
	\vspace{-1mm}\begin{equation}
		\bar{p}_\Theta(\theta)= \sum_{k=1}^{K}p_k\frac{1}{\sigma_{\theta}\sqrt{2\pi}}{\rm{e}}^{-\frac{(\theta-\theta_k)^2}{2\sigma_{\theta}^2}}. \label{cons: barptheta}	\vspace{-1mm}
	\end{equation}
	Specifically, $\bar{p}_\Theta(\theta)$ is the weighted sum of $K$ Gaussian PDFs, where each $k$-th Gaussian PDF is centered at mean $\theta_k$ with a small variance $\sigma_\theta^2$, and carries a weight of $p_k$. Note that as $\sigma_\theta^2$ decreases, $\bar{p}_\Theta(\theta)$ becomes increasingly similar to ${p}_\Theta(\theta)$. Moreover, with a sufficiently small $\sigma_\theta^2$, the probability for $\theta$ under (\ref{cons: barptheta}) to exceed the original $[-\pi,\pi)$ region is negligible. 
	
	Based on (\ref{cons: barptheta}), the Fisher information matrix (FIM) for the estimation of $\mv{\omega}$ consists of two parts as follows \cite{bib_Shen}:
	\vspace{-2mm}\begin{equation}
		\mv{J} = \mv{J}_{D} + \mv{J}_{P}.	\vspace{-2mm}
	\end{equation}
	The first part $\mv{J}_{D}\in \mathbb{C}^{3\times 3}$ represents the FIM extracted from the observed data in ${\mv{y}}_R$, which is given by
	\begin{align}
		\left\lbrack \mv{J}_{D} \right\rbrack_{ij}=-\mathbb{E}_{\mv{y}_R,\mv{\omega}}\left\lbrack \frac{\partial^{2}L_{\mv{y}_R}(\mv{\omega})}{\partial\omega_{i}\partial\omega_{j}} \right\rbrack,\ i,j\in \{1,2,3\},
	\end{align}
	with $L_{\mv{y}_R}(\mv{\omega})\!=\!-N_r{\rm{ln}}(\pi\sigma_R^2)\!-\!\frac{1}{\sigma_R^2}(\|\mv{y}_R\|^2\!+\!|\beta|^2\|\mv{M}(\theta)\mv{x}\|^2)+\frac{2}{\sigma_R^2}{\rm{Re}}\{\beta^*\mv{x}^H\mv{M}^H(\theta)\mv{y}_R\}$ being the log-likelihood function for the parameters in $\mv{\omega}$. $\mv{J}_D$ can be further derived as
	\begin{align}
		\mv{J}_{D} = \begin{bmatrix}
			J_{\theta\theta}&\mv{J}_{\theta\beta} \\ 
			\mv{J}_{\theta\beta}^H&  \mv{J}_{\beta\beta}
		\end{bmatrix},
	\end{align}
	where each block is given as
	\begin{align}
		&J_{\theta\theta}= \frac{2|\beta|^2}{\sigma_{R}^{2}}\int_{-\infty}^{\infty}\bar{p}_\Theta(\theta){{ \rm{tr}\left( {\dot{\mv{M}}^{\it{H}}\left( \theta \right)\dot{{\mv{M}}}\left( \theta \right)\mv{R}_{\mathit{x}}} \right)}}{\rm{d}}\theta,\\
		&\mv{J}_{\theta\beta}= 
		\frac{2}{\sigma_{R}^{2}}\int_{-\infty}^{\infty}\bar{p}_\Theta(\theta){\rm{tr}}\left(\dot{\mv{M}}^H(\theta)\mv{M}^H(\theta)\mv{R}_x\right)
		[{\beta_R,\beta_I}]{\rm{d}}\theta,\\
		&\mv{J}_{\beta\beta}= \frac{2}{\sigma_{R}^{2}}\int_{-\infty}^{\infty}\bar{p}_\Theta(\theta){\rm{tr}}\left( \mv{M}^{H}\left( \theta \right)\mv{M}\left( \theta \right)\mv{R}_{\mathit{x}} \right)\mv{I}_{2}{\rm{d}}\theta,
	\end{align}
	with $\dot{{\mv{M}}}(\theta)=\frac{\partial\mv{M}(\theta)}{\partial\theta}$ and $\mv{R}_x=\mathbb{E}[\mv{x}\mv{x}^H]=\mv{ww}^H+\sum_{k=1}^K\mv{v}_k\mv{v}_k^H$ denoting the transmit covariance matrix.
	
	The second part $\mv{J}_{P}\in \mathbb{C}^{3\times 3}$ represents the FIM extracted from the prior distribution information, which is given by
	\begin{align}
		\left\lbrack \mv{J}_{P} \right\rbrack_{ij} = - \mathbb{E}_{\mv{\omega}}\left\lbrack \frac{\partial^{2}{\ln p_{\mv{w}}}( \mv{\omega} )}{\partial\omega_{i}\partial\omega_{j}} \right\rbrack,\ i,j\in \{1,2,3\},
	\end{align}
	where $p_{\mv{w}}(\mv{\omega})$ denotes the PDF of $\mv{\omega}$. Note that since $\beta_R$ and $\beta_I$ are both deterministic variables, $\mv{J}_P$ only has a non-zero entry in the first column and first row, which is given by
	\begin{align} \label{cons: J_D intergral}
		\left\lbrack \mv{J}_{P} \right\rbrack_{\theta\theta} = -\int_{-\infty}^{+\infty}\frac{\partial^2\bar{p}_\Theta(\theta)}{\partial^2 \theta}d\theta+ \int_{-\infty}^{+\infty}\frac{\left(\frac{\partial  \bar{p}_\Theta(\theta)}{\partial \theta}\right)^2}{\bar{p}_\Theta(\theta)}d\theta. 
	\end{align}
	The first term in the right-hand side of (\ref{cons: J_D intergral}) can be derived as
	\begin{align}
		\!\!\!-\int_{-\infty}^{+\infty}\frac{\partial^2\bar{p}_\Theta(\theta)}{\partial^2 \theta}d\theta\!= \!\!\sum_{k=1}^{K}\frac{p_k(\theta-\theta_k)}{\sigma_\theta^3\sqrt{2\pi}}{\rm{e}}^{-\frac{(\theta-\theta_k)^2}{2\sigma_\theta^2}}\bigg|_{-\infty}^{+\infty}\!=\!0.
	\end{align}
	The second term can be derived as
	\begin{align}
		\int_{-\infty}^{+\infty}\frac{\Big(\frac{\partial  \bar{p}_\Theta(\theta)}{\partial \theta}\Big)^2}{\bar{p}_\Theta(\theta)}d\theta \! =&\!\! \int_{-\infty}^{+\infty}\frac{\sum_{k=1}^{K}\Big(p_k\frac{(\theta-\theta_k)}{\sigma_{\theta}^3\sqrt{2\pi}}{\rm{e}}^{-\frac{(\theta\!-\!\theta_k)^2}{2\sigma_{\theta}^2}}\Big)^2}{\bar{p}_\Theta(\theta)}d\theta  \nonumber\\
		-\epsilon
		= &\sum_{k=1}^{K}p_k\frac{1}{\sigma^2_\theta}-\epsilon= \frac{1}{\sigma^2_\theta}-\epsilon,
	\end{align}
	where $\epsilon\!\overset{\Delta}{=}\!\int_{-\infty}^\infty\sum\limits_{k=1}^{K}\!\sum\limits_{n=1 }^{K}f_{k}(\theta)f_{n}(\theta)\frac{(\theta_n-\theta_{k})^2}{\sigma_{\theta}^4}/(2\sum\limits_{k=1}^{K}f_k(\theta))d\theta$ with $f_k(\theta)\!\!\overset{\Delta}{=}\!\!\frac{p_k}{\sigma_{\theta}\sqrt{2\pi}}{\rm{e}}^{-\frac{(\theta\!-\!\theta_k)^2}{2\sigma_{\theta}^2}}$. Thus, we have $\left\lbrack \mv{J}_{P} \right\rbrack_{\theta\theta}\! =\! \frac{1}{\sigma^2_\theta}-\epsilon$. 
	
	Based on the above, the overall FIM $\mv{J}$ can be expressed as
	\begin{align}
		\mv{J} = \mv{J}_D+\mv{J}_P = \begin{bmatrix}
			J_{\theta\theta}+\frac{1}{\sigma^2_\theta}-\epsilon&\mv{J}_{\theta\beta} \\ 
			\mv{J}_{\theta\beta}^H &  \mv{J}_{\beta\beta}
		\end{bmatrix}.
	\end{align}
	The PCRB for the estimation MSE of ${\theta}$ denoted by ${\rm{PCRB}}_\theta$ is then given by the entry in the first column and first row of $\mv{J}^{-1}$, which can be expressed as
	\begin{align}\label{cons:pcrb}
		&{\rm{PCRB}}_\theta 
		= \left[J_{{\theta}{\theta}}+\frac{1}{\sigma^2_\theta}-\epsilon-\mv{J}_{{\theta}{\beta}}\mv{J}_{\beta\beta}^{-1}\mv{J}_{\theta\beta}^{H}\right]^{-1}\\ 
		=& \frac{\sigma_R^2{g}_1(\mv{R}_x)}{2|\beta|^2\left(\left({g}_2(\mv{R}_x)+\frac{\sigma_R^2}{2|\beta|^2}\left(\frac{1}{\sigma_\theta^2}-\epsilon\right)\right){g}_1(\mv{R}_x)-\left|{g}_3(\mv{R}_x)\right|^2\right)}.\nonumber
	\end{align}
	where ${g}_1(\mv{R}_x)=\int_{-\infty}^{\infty}\bar{p}_\Theta(\theta){\rm{tr}}({{\mv{M}}^{H}(\theta)\mv{M}(\theta )\mv{R}_{\mathit{x}}}){\rm{d}}\theta$; ${g}_2(\mv{R}_x) =  \int_{-\infty}^{\infty}\bar{p}_\Theta(\theta){\rm{tr}}({\dot{\mv{M}}^{\it{H}}( \theta )\dot{{\mv{M}}}(\theta)\mv{R}_{\mathit{x}}}){\rm{d}}\theta$; and ${g}_3(\mv{R}_x)= \int_{-\infty}^{\infty}\bar{p}_\Theta(\theta){\rm{tr}}({\dot{{\mv{M}}}^{H}(\theta)\mv{M}(\theta)\mv{R}_{\mathit{x}}}){\rm{d}}\theta$.
	
	Note that the PCRB in (\ref{cons:pcrb}) is a complicated function with respect to the transmit covariance matrix $\mv{R}_x$ and consequently the beamforming vectors $\mv{w}$ and $\mv{v}_k$'s, which is difficult to be theoretically analyzed or numerically examined. Motivated by this, we derive a more tractable upper bound of PCRB as follows. Specifically, we first re-express (\ref{cons:pcrb}) as
	\begin{align}
		\!\!\!\!{\rm{PCRB}}_\theta
		\!=\! \frac{\frac{\sigma_R^2}{2|\beta|^2}g_1(\mv{R}_x)}{g_4(\mv{R}_x)g_1(\mv{R}_x)\!+\!\frac{\sigma_R^2g_1(\mv{R}_x)}{2|\beta|^2}\left(\frac{1}{\sigma_\theta^2}\!-\!\epsilon\right)\!+\!g_5(\mv{R}_x)},\!\!\!\!
	\end{align}
	where ${g}_4(\mv{R}_x) = \int_{-\infty}^{\infty}\bar{p}_\Theta(\theta)\|\dot{\mv{b}}(\theta)\|^2\mv{a}^H(\theta)\mv{R}_x\mv{a}(\theta){\rm{d}}\theta$; ${g}_5(\mv{R}_x) =  \frac{1}{2}\int_{-\infty}^{\infty}\int_{-\infty}^{\infty}\|{\mv{b}}(\theta_p)\|^2\|{\mv{b}}(\theta_q)\|^2|\dot{\mv{a}}^H(\theta_p)\mv{R}_x{\mv{a}}(\theta_q)$ $-{\mv{a}}^H(\theta_p)\mv{R}_x\dot{\mv{a}}(\theta_q)|^2
	\bar{p}_\Theta(\theta_p)\bar{p}_\Theta(\theta_q)d\theta_pd\theta_q	\geq 0$. 
	By noting both ${g}_1(\mv{R}_x)$ and ${g}_5(\mv{R}_x)$ are non-negative, an upper bound of ${\rm{PCRB}}_\theta$ can be obtained as
	\begin{align}\label{Upp1}
		&{\rm{PCRB}}_\theta \leq  {\rm{PCRB}}^U_\theta\overset{\Delta}{=}\frac{1}
		{\frac{2|\beta|^2}{\sigma_R^2}{g_4}(\mv{R}_x)+\left(\frac{1}{\sigma_\theta^2}-\epsilon\right)}\nonumber\\
		=&	\frac{1}{\frac{2|\beta|^2}{\sigma_R^2}\left(\mv{w}^H\mv{Q}\mv{w}+\sum_{k=1}^K\mv{v}_k^H\mv{Q}\mv{v}_k\right)+\left(\frac{1}{\sigma^2_\theta}-\epsilon\right)}, 
	\end{align}
	where $\mv{Q}=\sum_{k=1}^K\int_{-\infty}^{+\infty}f_k(\theta)\|\dot{\mv{b}}(\theta)\|^2\mv{a}(\theta)\mv{a}(\theta)^H{\rm{d}}\theta$. Note that $\mv{Q}$ and $\epsilon$ in (\ref{Upp1}) still involve complicated integrals over the continuous $\theta$, for which an analytical expression is difficult to obtain. By leveraging the fact that we consider a small variance $\sigma_\theta^2$ in the Gaussian mixture model, we propose an approximation of (\ref{Upp1}) in \emph{closed form}, which will facilitate our optimization of the beamforming vectors in the next section. 
	
	\begin{proposition}\label{pro 1}
		With a small $\sigma_\theta^2$, an approximate expression for the PCRB upper bound ${\rm{PCRB}}^U_\theta$ is given by
		\begin{align} \label{cons: PCRB_value}
			\!\!\!\!{\rm{PCRB}}^U_\theta\!\approx\!\bar{\rm{PCRB}}^U_\theta\!\overset{\Delta}{=}\!{\frac{1}{\frac{2|\beta|^2}{\sigma_R^2}(\mv{w}^H\bar{\mv{Q}}\mv{w}\!+\!\sum_{k=1}^K\mv{v}_k^H\bar{\mv{Q}}\mv{v}_k)\!+\!\frac{1}{\sigma_\theta^2}}},
		\end{align}
		where $\!\!\bar{\mv{Q}}\!\!\overset{\Delta}{=}\!\!\rho_0\sum_{k=1}^K {p_k({\rm{cos}}(2\theta_k)\!+\!1)}\mv{a}(\theta_k)\mv{a}^{H}(\theta_k)\!\!$ with $\rho_0\overset{\Delta}{=}\sum_{n=1}^{N_r}\pi^2\Delta^2(n\!-\!1)^2$.
	\end{proposition}
	\begin{IEEEproof}
		Please refer to Appendix A.
	\end{IEEEproof}
	Notice that the approximate PCRB upper bound $\bar{\rm{PCRB}}^U_\theta$ has a closed-form expression which is an explicit function of the beamforming vectors $\mv{w}$ and $\mv{v}_k$'s, thus will be adopted as the performance metric for sensing the target's location. It is worth noting that the tightness of the PCRB upper bound ${\rm{PCRB}}^U_\theta$ with respect to the exact PCRB ${\rm{PCRB}}_\theta$ as well as the accuracy of its approximation $\bar{\rm{PCRB}}^U_\theta$ has been validated numerically with moderate values of $\sigma_\theta$ (e.g., $\sigma_\theta<10^{-2}$), for which the details are omitted due to limited space.
	
	Furthermore, note that $\bar{\rm{PCRB}}^U_\theta$ is a decreasing function of the amplitude of the overall reflection coefficient $\beta$ and consequently the amplitude of the target's unknown RCS coefficient $\alpha$. Thus, a global upper bound of the PCRB that holds for any value of $\alpha$ can be obtained by considering the minimum value of $|\alpha|$ denoted by $|\bar{\alpha}|=\min |\alpha|$ (which can be obtained \emph{a priori} by exploiting properties of the target), and replacing $\beta$ in (\ref{cons: PCRB_value}) with the corresponding $|\bar{\beta}|=\frac{\beta_0}{r^2}|\bar{\alpha}|=\min |\beta|$.
	
	\section{Problem Formulation}
	In this paper, our objective is to optimize the transmit beamforming vectors $\mv{w}$ and $\mv{v}_k$'s to maximize the worst-case secrecy rate among all possible eavesdropper locations, while ensuring the sensing PCRB for the eavesdropping target is always below a given threshold $\Gamma$. Motivated by the tractability of the approximate PCRB upper bound $\bar{\rm{PCRB}}^U_\theta$, we aim to achieve this goal by ensuring that $\bar{\rm{PCRB}}^U_\theta$ corresponding to the minimum RCS amplitude $|\bar{\alpha}|$ is no larger than $\Gamma$. Thus, the optimization problem is formulated as follows.
	\begin{align}  
		\!\!\!\!\!({\rm{P1}})\!\ {\underset{\mv{w},\{\!\mv{v}_k\!\}}{\rm max}}\!\ {\underset{k}{\rm{min}}}\!\ &\log_2(1\!+\!\mathrm{SINR})\!-\!\log_2(1\!+\!\mathrm{SINR}_E(\theta_k\!)\!)\! \\ 
		{\rm{s.t.}}\   &\!{\left\| \mv{w} \right\|^{2} + \sum_{k=1}^K\left\| \mv{v}_k \right\|^{2} \leq P} \\ 
		&\!\frac{1}{\frac{2|\bar{\beta}|^2}{\sigma_R^2}(\mv{w}^H\bar{\mv{Q}}
			\mv{w}\!+\!\sum_{k=1}^K\!\!\mv{v}_k^H\bar{\mv{Q}}\mv{v}_k\!)\!+\!\frac{1}{\sigma^2_\theta}}\!\leq\! \Gamma.\!\!\!\!\label{P1c3}
	\end{align}
	Note that the objective function in (P1) involves logarithm functions of fractional quadratic functions, and can be shown to be non-concave. Moreover, the constraint in (\ref{P1c3}) is also non-convex since $\bar{\mv{Q}}$ can be observed to be a positive semi-definite (PSD) matrix. Therefore, (P1) is a non-convex problem. 
	
	Moreover, it is worth noting that in order to maximize the secrecy rate, $\mv{w}$ should be designed such that the received power of the information beam at each possible target location is as small as possible; on the other hand, to minimize the approximate PCRB upper bound, $\mv{w}$ should be designed to maximize $\mv{w}^H\bar{\mv{Q}}\mv{w}$, which is proportional to the weighted summation of the received information beam powers among all possible target locations. Therefore, there exists a non-trivial trade-off between the secrecy performance and sensing performance in the secure ISAC system. Furthermore, note that both the secrecy rate and the PCRB are critically dependent on the distribution of $\theta$, which makes the problem   more challenging. In the following, we derive the optimal \hbox{solution to (P1).}

	\section{Optimal Solution to (P1)}
	\subsection{Equivalent Problem Reformulation}
	First, note that one key difficulty in (P1) lies in the fractional expressions of the SINR. To deal with this issue, we introduce an auxiliary variable $\gamma$ to characterize the SINR constraint at each possible eavesdropper (target) location. Based on this, according to \cite{bib13}, \cite{bib15}, it can be proved that there always exists a $\gamma>0$ at all possible eavesdropper locations such that problem (P1.1) below has the same optimal solution with (P1).
	\begin{align} \label{con: sub1}
		\!\!\!\!	({\rm{P1.1}}){\underset{\mv{w},\{\mv{v}_k\}}{\rm max}~} & {{{\frac{|{\mv{h}^{H}\mv{w}}|^{2}}{\sum_{k=1}^K|\mv{h}^{H}\mv{v}_k|^{2} + \sigma^{2}}} }} \\ 
		{\rm{s.t.}} \ &  \frac{|\mv{a}(\theta_k)^{H}\mv{w}|^{2}}{\sum_{k=1}^K|\mv{a}(\theta_k)^{H}\mv{v}_k|^{2}\!+\! \sigma_E^2r^2/\beta_0}\leq \gamma,\ \forall k\!\\ 
		& {\left\| \mv{w} \right\|^{2} + \sum_{k=1}^K\|\mv{v}_k\|^{2} \leq P} \\ 
		& \mv{w}^H\bar{\mv{Q}}
		\mv{w}\!+\!\!\sum_{k=1}^K\mv{v}_k^H\bar{\mv{Q}}\mv{v}_k \!\geq\! \frac{\sigma_R^2}{2|\bar{\beta}|^2}\left(\frac{1}{\Gamma}\!-\!\frac{1}{\sigma^2_\theta}\right)\!.\!\!\!
	\end{align}
	Moreover, denote $f(\gamma)$ as the optimal value of (P1.1) with a given $\gamma>0$. Then, the following problem can be shown to have the same optimal value as (P1) \cite{bib13}:
	\begin{align} \label{con: sub2}
		({\rm{P1.2}}) \qquad{\underset{\gamma > 0}{{\rm{max}}}~}  \qquad \log_2((1 + f(\gamma))/(1 + \gamma)).
	\end{align}
	Therefore, the optimal solution to (P1) can be obtained via one-dimensional search of $\gamma>0$ in (P1.2) based on the values of $f(\gamma)$. Thus, our remaining task is to obtain the optimal solution to (P1.1). 
	\subsection{Optimal Solution to (P1.1)}
	Motivated by the quadratic functions involved in (P1.1), we consider an SDR based approach. Let $\mv{H}\overset{\Delta}{=}\mv{h}\mv{h}^{H}$, $\mv{W}\overset{\Delta}{=}\mv{w}\mv{w}^{H}$, $\mv{V}\overset{\Delta}{=}\sum_{k=1}^K\mv{v}_k\mv{v}_k^{H}$, and $\mv{A}_{k} \overset{\Delta}{=}\mv{a}(\theta_k)\mv{a}^H(\theta_k),\forall k$. Then, (P1.1) can be equivalently expressed as the following problem with an additional constraint of $\mathrm{rank}(\mv{W})=1$:
	\begin{align}\label{con: Problem sub1SDR1} 
\hspace{-2mm}\!\!\!\!\mathrm{(P1.1R)}\	{\underset{\mv{W},\mv{V}}{{\rm{max}}}~} \  & { \frac{{\rm{tr}}\left( {\mv{H}\mv{W}} \right)}{{\rm{tr}}\left( \mv{H}\mv{V} \right) + \sigma^{2}}}  \\ 
		{\rm{s.t.}} \ & {\rm tr}\left( {\mv{A}_{k}\mv{W}} \right)\! \leq\! \gamma\left( {\rm{tr}}\left({\mv{A}_k\mv{V}}\right)\! +\! \frac{\sigma_E^2r^2}{\beta_0} \right),\forall k\label{P1.1c1}\\ 
		& {\rm{tr}}\left( \mv{W} \right) + {\rm{tr}}\left( \mv{V} \right) \leq P \\ 
		& {\rm{tr}}((\mv{W}+\mv{V})\bar{\mv{Q}})\geq \frac{\sigma_R^2}{2|\bar{\beta}|^2}\left(\frac{1}{\Gamma}-\frac{1}{\sigma^2_\theta}\right)\\ 
		& \mv{W} \succeq \mv{0},\ \mv{V} \succeq \mv{0}.\label{P1.1c5}
	\end{align}	
	The objective function of (P1.1R) is non-concave. However, we can leverage the Charnes-Cooper transformation \cite{bib14} to transform (P1.1R) into an equivalent convex problem as:
	\begin{align} \label{con: Problem sub1SDR1_2}
\hspace{-2mm}({\rm{P2.1R}}){\underset{\mv{W},\mv{V},{t}}{{\rm{max}}}} \ & {\rm tr}\left( {\mv{H}\mv{W}} \right)  \\ 
		{\rm{s.t.}} \ &\! {\rm tr}\left( {\mv{A}_{k}\mv{W}} \right)\! \leq\! \gamma\left(\!{\rm{tr}}\left({\mv{A}_k\mv{V}}\right)\! +\! \frac{t\sigma_E^2r^2}{\beta_0}\!\right),\!\forall k\label{P2.1c1}\\ 
 		& {\rm tr}\left( {\mv{H}\mv{V}} \right) + t\sigma^{2} = 1\label{P2.1c2}\\
		& {\rm{tr}}\left( \mv{W} \right) + {\rm{tr}}\left( \mv{V} \right) \leq tP\label{P2.1c3} \\ 
		&  {\rm{tr}}((\mv{W}+\mv{V})\bar{\mv{Q}})\geq \frac{t\sigma_R^2}{2|\bar{\beta}|^2}\left(\frac{1}{\Gamma}-\frac{1}{\sigma^2_\theta}\right) \label{P2.1c4}\\  
		&  (\ref{P1.1c5}), t > 0.\label{P2.1c5}
	\end{align} 
	
	Since (P2.1R) is a convex optimization problem, its optimal solution can be obtained efficiently via interior-point method or CVX. Moreover, the duality gap is equal to zero. Let $\left\{\beta_{k}\right\}$, $\lambda$, $\rho$, and $\psi$ denote the dual variables associated with the constraint(s) in (\ref{P2.1c1}), (\ref{P2.1c2}), (\ref{P2.1c3}), and (\ref{P2.1c4}), respectively. The Lagrangian of (P2.1R) is given by
	\begin{align}\label{con: L}
		\!\!	\mathcal{L}( {\mv{W},\mv{V},t,\{\beta_{k}\},\lambda,\rho,\psi} )\! =\! {\rm tr}( {\mv{S}\mv{W}} ) \!+\! {\rm tr}( {\mv{B}\mv{V}} ) \!+\! \xi t \!+\! \lambda,
	\end{align}
	where $\mv{S}= \mv{H} - {\sum_{k = 1}^{K}{\beta_{k}\mv{A}_{k} + \psi\bar{\mv{Q}} - \rho\mv{I}_{N_t}}}$; $\mv{B} = - \lambda\mv{H} + {\gamma\sum_{k = 1}^{K}{\beta_{k}\mv{A}_{k} }}+\psi\bar{\mv{Q}} - \rho\mv{I}_{N_t}$; and $\xi = - \lambda\sigma^{2} + {\gamma\frac{\sigma_E^2r^2}{\beta_0}\sum_{k = 1}^{K}{\beta_{k}}} + \rho P - \psi\frac{\sigma_R^2}{2|\bar{\beta}|^2}(\frac{1}{\Gamma}-\frac{1}{\sigma^2_\theta})$.
	
	Let $\lambda^*$, $\{\beta_k^*\geq 0\}$, $\rho^*\geq 0$, and $\psi^*\geq 0$ denote the optimal dual variables to (P2.1R) Then, we have the following lemma.
	
	\begin{lemma}\label{lemma 1}
		The optimal dual solution satisfies $\lambda^*>0$ and $\rho^*>0$ when $\gamma>0$.
	\end{lemma}
	\begin{IEEEproof}
		Please refer to Appendix B.	
	\end{IEEEproof}
	
	Since $\rho^*>0$, the constraint in (\ref{P2.1c3}) must be satisfied with equality by the optimal solution to (P2.1R) due to the complementary slackness. Define $\mv{D}^* = - \lambda^{*}\mv{H} - {\sum_{k = 1}^{K}{\beta_{k}^{*}\mv{A}_{k}}} + \psi^{*}{{\mv{A}_{k}}} - \rho^{*}\mv{I}_{N_t}$ 	with $l = {\rm{rank}}(\mv{D}^*)$. The orthogonal basis of the null space of $\mv{D}^*$ can be represented as $\mv{Z}\in \mathbb{C}^{N_{t} \times {({N_{t} - l})}}$, where $\mv{z}_{1,n}$ denotes the $n$-th column of $\mv{Z}$; if $l=N_t$, $\mv{Z} = \mv{0}$. 
	Then, we have the following proposition for (P2.1R).
	
	\begin{proposition}\label{pro 2}
		For (P2.1R), the optimal $\mv{V}^*$ satisfies ${\rm{rank}}(\mv{V}^*)\leq {\rm{min}}(K, N_t)$. The optimal $\mv{W}^*$ can be \hbox{written as} 
		\begin{align}
			\mv{W}^*  = \sum_{n=1}^{N_t-l}{a_n\mv{z}_{1,n}\mv{z}_{1,n}^H+b\mv{r}\mv{r}^H}, \ \label{con: W}
		\end{align}
		where $a_n\geq0$ for $\forall n$, $b>0$, and $\mv{r}\in \mathbb{C}^{N_{t} \times 1}$ satisfies $\mv{r}^H\mv{Z} = \mv{0}$. If ${\rm{rank}}(\mv{W}^*) >1$, the following set of the solution with a rank-one solution of $\mv{W}$ can be constructed which achieves the same optimal value of (P2.1R):
		\begin{align}
			&{\overline{\mv{W}}}^* = b\mv{r}\mv{r}^H \label{W}\\
			&{\overline{\mv{V}}}^* = \mv{V} + \sum_{n=1}^{N_t-l}a_n\mv{z}_{1,n}\mv{z}_{1,n}^H \label{V}\\
			&{\overline{t}^*} = t^*.\label{t}
		\end{align}
	\end{proposition}
	\begin{IEEEproof}
		Please refer to Appendix C.
	\end{IEEEproof}
	
	To summarize, if the optimal solution $({\mv{W}}^*, {\mv{V}}^*, t^*)$ of (P2.1R) satisfies ${\rm{rank}}({\mv{W}}^*) = 1$, we can get an optimal solution $({\mv{W}}^*/t^*, {\mv{V}}^*/t^*)$ for (P1.1R) and consequently (P1.1). Otherwise, we can reconstruct 
	$({\overline{\mv{W}}}^*, {\overline{\mv{V}}}^*, {\overline{t}}^*)$ with ${\rm{rank}}(\overline{\mv{W}}^*) = 1$ according to (\ref{W}), (\ref{V}), and (\ref{t}), and $(\overline{\mv{W}}^*/t^*, \overline{\mv{V}}^*/t^*)$ will be the optimal solution for (P1.1).

	\section{Numerical Results}
	In this section, we provide numerical results to evaluate the performance of the proposed transmit beamforming design for secure ISAC. We assume the BS is equipped with $N_t = 8$ transmit antennas and $N_r = 10$ receive antennas, where the antenna spacing is half of the wavelength (i.e., $\Delta=\frac{1}{2}$). 	We consider $K=4$ possible target (eavesdropper) locations, with $\theta_1=-55^{\circ},\theta_2=-35^{\circ},\theta_3=65^{\circ}$, $\theta_4=45^{\circ}$; $p_1=0.2,p_2=0.3,p_3=0.1$, and $p_4=0.4$. The transmit power is set as $P=20$ dBm. The path loss of the BS-target channel is $10$ dB. The lower bound of the target's RCS is set as $|\bar{\alpha}|=0.0071$. The average receiver noise power is set as $\sigma_R^2=\sigma^2_E = \sigma^2 = -60~{\rm{dBm}}$. The BS-user channel follows an LoS model, where the user's angle with respect to the BS reference point is $-10^\circ$.
	
\begin{figure}
		\centering
		\includegraphics[scale=0.35]{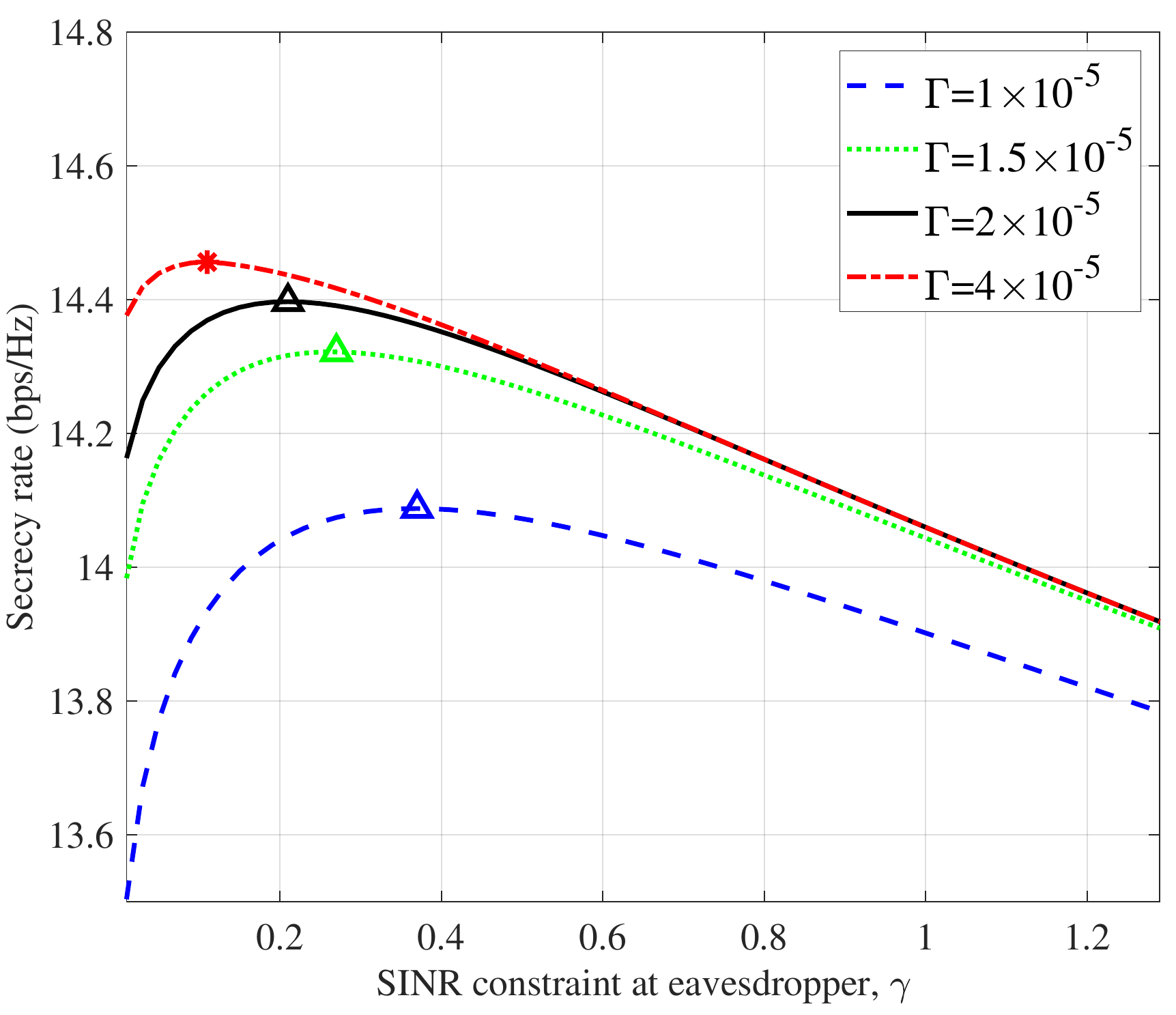}
		\vspace{-3mm}
		\caption{Secrecy rate versus the SINR constraint at the eavesdropper, $\gamma$.}
		\label{Fig2}
		\vspace{-3mm}
	\end{figure}

	First, in Fig. \ref{Fig2}, we illustrate the secrecy rate $\log_2((1+f(\gamma))/(1+\gamma))$ versus the SINR constraint at the eavesdropper, $\gamma$, for different sensing accuracy thresholds $\Gamma$'s. We set the path loss of the user as $30$ dB. It is observed that for every value of $\Gamma$, the secrecy rate first increases and then decreases with $\gamma$, and there exists a unique optimal solution of $\gamma$. Moreover, as the sensing accuracy constraint becomes more stringent (with a smaller $\Gamma$), it can be observed that the maximized secrecy rate decreases, which demonstrates the trade-off between secrecy and sensing.
	
\begin{figure}
		\centering
		\includegraphics[scale=0.35]{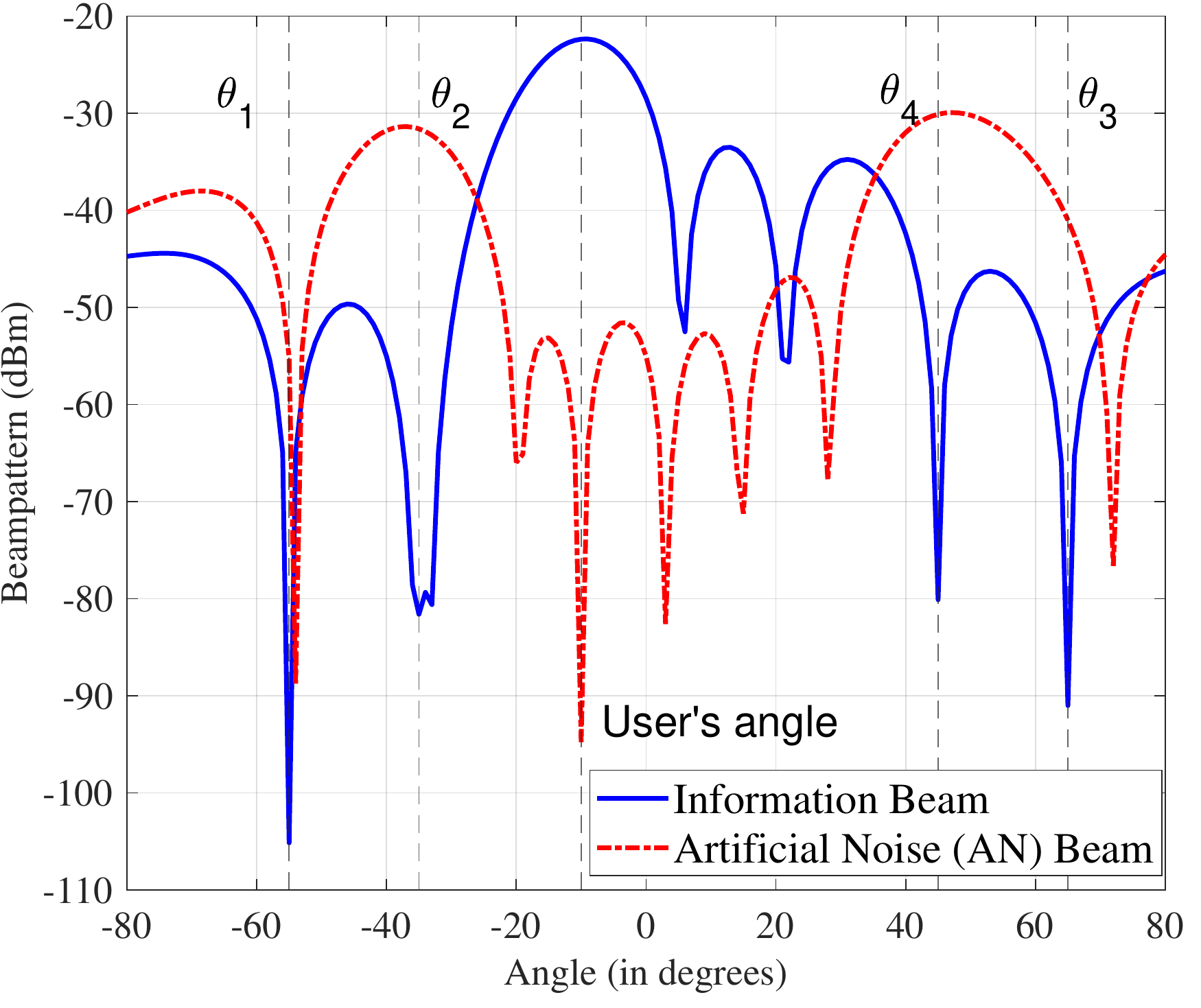}
		\vspace{-4mm}
		\caption{Beampattern versus angle.}
		\vspace{-8mm}
		\label{Fig3}
	\end{figure}

	Next, we consider $\Gamma=2.68\times 10^{-5}$ and illustrate in Fig. \ref{Fig3} the beampattern over different angles at distance $r$. The path loss of the user is set as $60~\rm{dB}$. It is observed that at the user's angle, the information beam is much stronger than the AN beam; whereas at the possible eavesdropper's locations, the information beam's power reaches its local minimum, but the AN beam power is generally very strong. This thus validates the effectiveness of our proposed beamforming design for enhancing the communication secrecy and sensing accuracy. 
	
\begin{figure}
		\centering
		\includegraphics[scale=0.35]{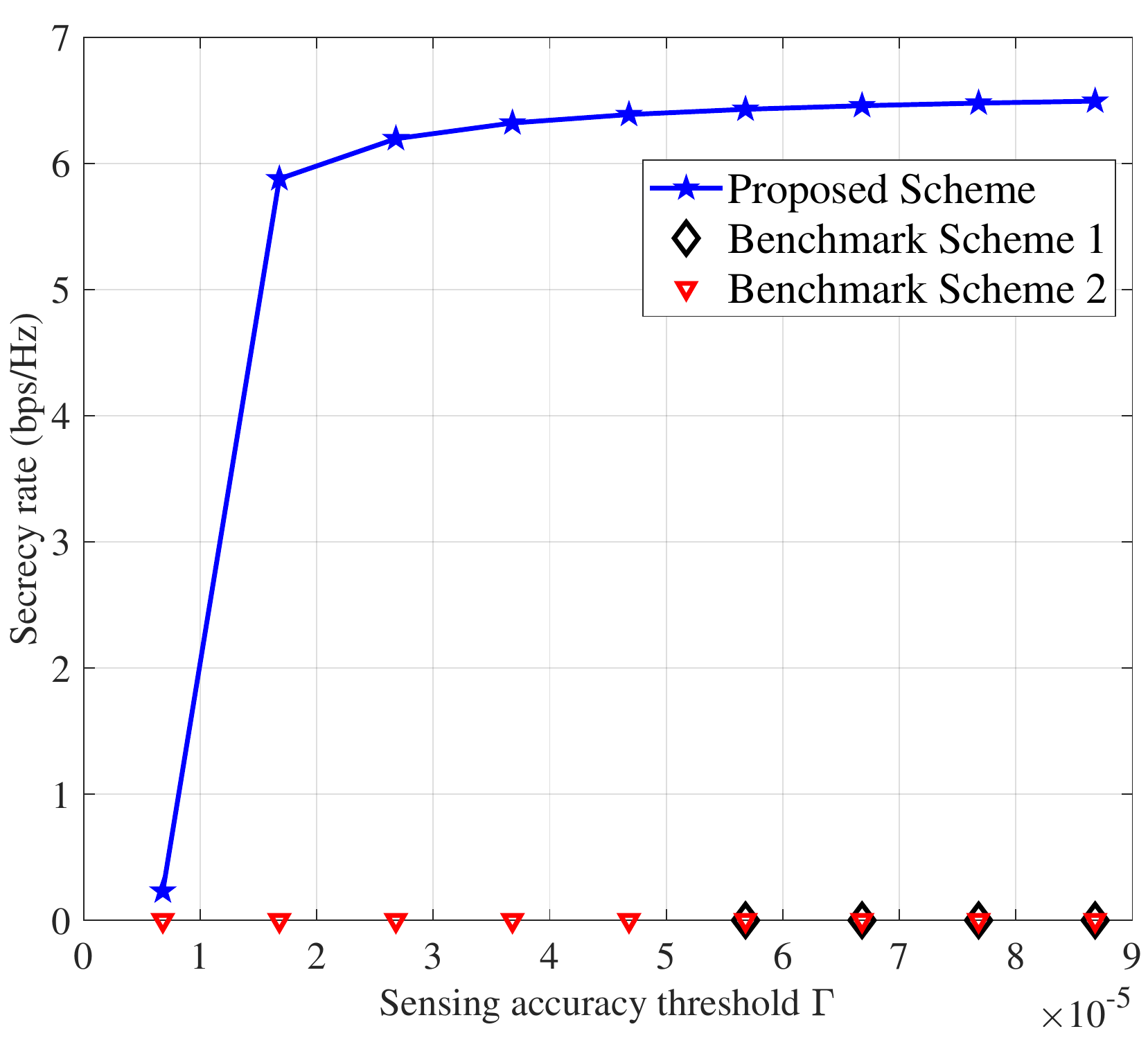}
		\vspace{-4mm}
		\caption{Secrecy rate versus sensing accuracy threshold for different schemes.}\label{Fig4}
		\vspace{-7mm}
	\end{figure}

	Finally, we compare the performance of our proposed transmit beamforming scheme with two benchmark schemes.
	\begin{itemize}
		\item {\bf{Benchmark 1}}: Simple maximum ratio transmission (MRT) beamforming with only an information beamforming vector $\mv{w}=\sqrt{P}\mv{h}/\|\mv{h}\|$, which is designed to maximize the user's achievable rate without consideration of the secrecy and sensing performance.
		\item {\bf{Benchmark 2}}: Beamforming without AN, where $\mv{w}$ is obtained by solving (P1) with $\mv{v}_k=0,\forall k$.
	\end{itemize}
	In Fig. \ref{Fig4}, we show the secrecy rate versus the sensing accuracy threshold $\Gamma$. It is observed that with our proposed scheme, the secrecy rate increases as the sensing accuracy constraint becomes less stringent, which further demonstrates the secrecy-sensing trade-off. On the other hand, Benchmark Scheme 1 is observed to be infeasible for the first five samples of $\Gamma$, due to its lack of consideration of the sensing performance; while both Benchmark Scheme 1 and Benchmark Scheme 2 fail to achieve a non-zero secrecy rate, due to the incapability of controlling the information leakage without the AN beam. 
	
	\section{Conclusions}
	This paper investigated a secure ISAC system where the sensing target may potentially serve as an eavesdropper. The location of the target is unknown and random, while its distribution is available for exploitation. First, to characterize the sensing performance, we derived the PCRB of MSE exploiting prior distribution, and further proposed a novel approximate PCRB upper bound with a closed-form expression. Then, we considered an AN-based transmit beamforming structure, and formulated the beamforming optimization problem to maximize the minimum secrecy rate among all possible target (eavesdropper) locations, under a sensing accuracy constraint represented by an upper bound on the PCRB. Although the formulated problem is non-convex, we adopted the SDR technique to obtain the optimal beamforming solution. Numerical results revealed that our proposed scheme outperforms various benchmark schemes in terms of both secrecy and sensing.
	
		\appendices
		\section{Proof of Proposition 1}
		First, we denote $\mv{S}(\theta_k)=\int_{-\infty}^{+\infty}f_k(\theta)\|\dot{\mv{b}}(\theta)\|^2\mv{a}(\theta)\mv{a}(\theta)^H{\rm{d}}\theta$, which yields $\mv{Q}=\sum_{k=1}^{K}\mv{S}(\theta_k)$. Based on the definition of $\mv{a}(\theta)$ and $\mv{b}(\theta)$, $\mv{S}(\theta_k)$ can be further expressed as
		\begin{align}\label{cons: S}
			&\mv{S}(\theta_k)=\left(2\sum_{n=1}^{N_r}\pi^2\Delta^2(n-1)^2\right)\int_{-\infty}^{+\infty} f_k(\theta){{\rm{cos}}^2(\theta)}\times\\
			&{\begin{bmatrix}
					1&\!\!\!\!e^{-j\pi2\Delta\sin\theta}&\!\!\!\!... &\!\!\!\!e^{-j\pi2(N_t-1)\Delta\sin\theta}\\ 
					e^{j\pi2\Delta\sin\theta} &\!\!\!\! 1 & \!\!\!\!... &\!\!\!\! e^{-j\pi2(N_t-2)\Delta\sin\theta}\\ 
					\vdots & \ddots&  &\vdots\\ 
					e^{j\pi2(N_t-1)\Delta\sin\theta}&\!\!\!\!e^{j\pi2(N_t-2)\Delta\sin\theta} &\!\!\!\!...&\!\!\!\! 1
			\end{bmatrix}}{\rm{d}}\theta.\nonumber
		\end{align}
		Denote $\mu_1 = \frac{p_k}{\sqrt{2\pi}}\left(2\sum_{n=1}^{N_r}\pi^2\Delta^2(n-1)^2\right)$, ${\mu}_2 = -2\pi\Delta$, and $t\!=\!\theta\!-\!\theta_k$. Then, $[\mv{S}(\theta_k)]_{1,2}$ can be further simplified as:
		\begin{align}
			&[\mv{S}(\theta_k)]_{1,2}=\mu_1\frac{1}{\sigma_\theta}\int_{-\infty}^{+\infty}{e}^{{-\frac{(\theta-\theta_k)^2}{2\sigma_{\theta}^2}}}{\rm{cos}}^2(\theta)e^{\mu_2 j{\rm{sin}}(\theta)} {\rm{d}}\theta\\
			&=\frac{\mu_1}{2\sigma_\theta}\int_{-\infty}^{\infty}{e}^{{-\frac{t^2}{2\sigma_{\theta}^2}}}e^{\mu_2 j(\sin\theta_k\cos t+\cos\theta_k\sin t)}\nonumber\\
			&\qquad\quad\times(\cos (2t)\cos (2\theta_k)-\sin (2t)\sin (2\theta_k)+1 ) dt\nonumber\\
			&\overset{(a)}{\approx}\![\tilde{\mv{S}}(\theta_k)]_{1,2}\!=\!\frac{\mu_1}{2\sigma_\theta}\int_{-\infty}^{+\infty}{e}^{{-\frac{t^2}{2\sigma_{\theta}^2}}}(\alpha_0\!+\!\alpha_1t\!+\!\alpha_2t^2\!+\!o(t^3))dt,\nonumber
		\end{align}
		where $\alpha_0 = e^{\mu_2j{\rm{sin}}(\theta_k)}({\rm{cos}}(2\theta_k)+1)$; $\alpha_1 = -2{\rm{sin}}(2\theta_k)-j\mu_2(1+{\rm{cos}}(2\theta_k)){\rm{sin}}(\theta_k)$; and $	\alpha_2 = -2{\rm{cos}}(2\theta_k)-\frac{1}{2}\mu_2(j{\rm{sin}}(\theta_k)+{\rm{cos}}(\theta_k))({\rm{cos}}(\theta_k)+1)-2j\mu_2{\rm{sin}}(2\theta_k){\rm{cos}}(\theta_k)$. Note that $(a)$ is derived by taking the Maclaurin series of ${\rm{cos}}(2t)$, ${\rm{sin}}(2t)$, $e^{\mu_2j{\rm{sin}}(\theta_k){\rm{cos}}(t)}$ and $e^{\mu_2j{\rm{cos}}(\theta_k){\rm{sin}}(t)}$ and noting $\sigma^2_\theta$ is a small value. Notice that ${e}^{{-\frac{t^2}{2\sigma_{\theta}^2}}}t$ and ${e}^{{-\frac{t^2}{2\sigma_{\theta}^2}}}t^3$ are odd functions. Moreover, $
		\int_{-\infty}^{+\infty}{e}^{{-\frac{t^2}{2\sigma_{\theta}^2}}}{\rm{d}}t\! =\! {{\sqrt{2\pi}}}\sigma_\theta $, and $\int_{-\infty}^{+\infty}{e}^{{-\frac{t^2}{2\sigma_{\theta}^2}}}t^2{\rm{d}}t\!=\!\sqrt{2\pi}\sigma_\theta^3$. Thus, we have
		$[\tilde{\mv{S}}(\theta_k)]_{1,2}=\mu_1({\rm{cos}}(2\theta_k)+1){\sqrt{\frac{\pi}{2}}}e^{\mu_2j{\rm{sin}}(\theta_k)}$. Similarly, other entries in $\mv{S}(\theta_k)$ can be approximated in the same manner, which yields
		\begin{align}\label{cons: S_value}
			\tilde{\mv{S}}(\theta_k)= \mu_1\sqrt{\pi/2}({\rm{cos}}(2\theta_k)+1)\mv{a}(\theta_k)\mv{a}^H(\theta_k).
		\end{align}
		
		Secondly, due to the small value of $\sigma_\theta^2$, the non-zero values of $\partial  \bar{p}_\Theta(\theta)/(\partial \theta)$ will only occur in the close vicinity of $\theta_k$'s. Thus, we have $\int_{-\infty}^{+\infty}\frac{\Big(\frac{\partial  \bar{p}_\Theta(\theta)}{\partial \theta}\Big)^2}{\bar{p}_\Theta(\theta)}d\theta\approx \frac{1}{\sigma_\theta^2}$, i.e., $\epsilon\approx 0$. Based on this and (\ref{cons: S_value}), Proposition \ref{pro 1} is proved.

\section{Proof of Lemma 1}
First, we prove that $\lambda^* > 0$ when $\gamma>0$.
According to (\ref{con: L}), the dual problem of (P2.1R) can be given by
\begin{align}
	\begin{aligned}
		{\underset{\{\beta_k\},\lambda,\rho, \psi}{{\rm{min}}}~} \ \qquad & \lambda\\
		\rm{s.t.} \ \qquad \quad & {\mv{S}\preceq \mv{0}, \qquad \mv{B}\preceq \mv{0}, \qquad \xi \leq 0}\\
		& \{\beta_k \}\geq 0, \quad \forall k\\
		& \psi\geq0, ~~\rho\geq 0. \label{con: LPlambda}
	\end{aligned}
\end{align}
To ensure that the Lagrangian problem in (\ref{con: L}) is bounded so that the dual function exists, it follows that
\begin{align}
	\mv{S}^*\preceq \mv{0}, \qquad \mv{B}^*\preceq \mv{0}, \qquad \xi^* \leq 0 \ \label{con:proof1}
\end{align}
which are given by substituting the optimal dual solution $\lambda^*$.
Based on strong duality, the duality gap is zero, thus $\lambda^*$ is equal to the optimal value of (P2.1R). Hence, we have $\lambda^*>0$.

Next, we demonstrate $\rho^*>0$ by contradiction. We have ${\bar{\mv{Q}}} = \sum_{k=1}^{K}w_k\mv{A}_k$, where $w_k=\rho_0p_k({\rm{cos}}(2\theta_k)+1)\geq 0,~\forall k$. Assume $\phi = \{k|(\beta_k^*)^2 + (\psi^*w_k)^2>0, k=1,\dots,K\}$. Then, we prove that $\rho^*\neq 0$ by discussing the following two cases with the assumption that $\rho^*= 0$.

\begin{itemize}
	\item Case 1:  For $\phi=\emptyset$, we have $\mv{S}^* =\mv{H}\succeq \mv{0}$, which contradicts with (\ref{con:proof1}). We can get $\rho^*>0$.
	\item Case 2: For $\phi\neq\emptyset$, we have $\mv{B}^* =-\lambda^*\mv{H}+{\sum_{k \in \phi}{\beta_{k}^{*}\gamma\mv{A}_{k}}} + \psi^*{\bar{\mv{Q}}} = -\lambda^*\mv{H}+{\sum_{k \in \phi}{\beta_{k}^{*}\gamma\mv{A}_{k}}} + \sum_{k=1}^{K}\psi^*w_k\mv{A}_k$. Then, we $\sum_{k \in \phi}\left(\beta_{k}^{*}\gamma+\psi^*w_k\right)\mv{A}_k\succeq \mv{0}$ and $\lambda^*\geq0$. To ensure $\mv{B}^*\preceq \mv{0}$, it requires that any vector that lies in the null space of $\mv{H}$ must be in the null space of $\mv{A}_k, \forall k\in \phi$. However, in our scenario, the channel $\mv{h}$ and ${\mv{a}}_k$ for $\forall k$ is linearly independent. Thus, we can get $\rho^*>0$.
\end{itemize}

By combining the aforementioned cases, we can conclude that $\rho^*>0$. Therefore, Lemma \ref{lemma 1} is proven.

\section{Proof of Proposition 2}

The optimal solutions for (P2.1R) satisfy the Karush-kuhn-Tucker (KKT) conditions which can be expressed as
\begin{align}
	\mv{S}^*\mv{W}^* = \mv{0}\qquad \mv{B}^*\mv{V}^* = \mv{0}. \ \label{con: KKTs}
\end{align}

1) We prove that ${\rm{rank}}(\mv{V}^*)\leq{\rm{min}}(K, N_t)$.
If $K \geq N_t$, ${\rm{rank}}(\mv{V}^*)\leq N_t = {\rm{min}}(K, N_t)$. Next, we introduce the following Lemma \ref{lemma 2}.

\begin{lemma}\label{lemma 2}
Let $\mv{Y}$ and $\mv{X}$ be two matrices of the same dimension. It holds that ${\rm{rank}}(\mv{Y}+\mv{X})\geq {\rm{rank}}(\mv{Y})-{\rm{rank}}(\mv{X})$.
\end{lemma}

{\it{Proof:}} If $\mv{Y}$ and $\mv{X}$ are of the same dimension, ${\rm{rank}}(\mv{Y})+{\rm{rank}}(\mv{X})\geq{\rm{\mv{Y}}+\mv{X}}$. Thus, due to ${\rm{rank}}(\mv{X})={\rm{rank}}(-\mv{X})$, we have ${\rm{rank}}(\mv{Y}+\mv{X})+{\rm{rank}}(-\mv{X})\geq{\rm{rank}}(\mv{Y})$. 

We define $\mv{C}^* = -\lambda^*\mv{H}-\rho^*\mv{I}$. Due to $\lambda^*>0$ and $\rho^*>0$, we have $\mv{C}^*\prec\mv{0}$, it follows that ${\rm{rank}}(\mv{C}^*) = N_t$. Therefore, $\mv{B}^* = \mv{C}^* +\sum_{k=1}^{K}\left( \beta_k^*\gamma+\psi^* w_k\right){\mv{A}}_k$. Due to ${\rm{rank}}\left(\sum_{k=1}^{K} \left(\beta_k^*\gamma+\psi^* w_k\right){\mv{A}} _k\right) \leq K$ , based on Lemma \ref{lemma 2}, we have
\begin{align}
\hspace{-2mm}{\rm{rank}}(\mv{B}^*) &\geq{\rm{rank}}(\mv{C}^*)\!-\!{\rm{rank}}\left(\sum_{k=1}^{K} \!\left(\beta_k^*\gamma+\psi^* w_k\right){\mv{A}} _k\!\right) \\ \nonumber
	&\geq N_t - K.
\end{align}
By combining the two cases mentioned above, namely $K\geq N_t$ and $K <N_t$, we can deduce that ${\rm{rank}}(\mv{V}^*)\leq{\rm{min}}(K, N_t)$.

2) We prove that optimal solution $\mv{W}^*$ can be written as (44) from two cases as follows.
\begin{itemize}
	\item Case 1: ${\rm{rank}}(\mv{D}^*)= l = N_t$, where $\mv{D}^* = \mv{B}^*-\sum_{k=1}^{K}(1+\gamma)\beta_k^*\mv{A}_k$. Then, we have
	\begin{align}
		\mv{S}^* = \mv{D}^*+(1+\lambda^*)\mv{H}.  \label{con: S}
	\end{align}
	According to $\mv{S}^*\mv{W}^*=0$ and {\it{Lemma 2}}, we can conclude that ${\rm{rank}}(\mv{S}^*)\geq N_t - 1$. If ${\rm{rank}}(\mv{S}^*)=N_t$, it follows that $\mv{W}^*={\mv{0}}$. It is not true for the optimal solution. Then, we conclude that ${\rm{rank}}(\mv{W}^*) = b\mv{r}\mv{r}^H$.
	\item Case 2: $l < N_t$. We have
	\begin{align}
\hspace{-2mm}\mv{z}_{1,n}^H\mv{S}^*\mv{z}_{1,n}= (1+\lambda^*)|\mv{h}^H\mv{z}_{1,n}|^2, 1\leq n\leq N_t-l, \label{con: norm0}
	\end{align}
	by taking (\ref{con: S}). Due to $\mv{S}^*\preceq \mv{0}$ and $(1+\lambda^*)>0$, we can get $|\mv{h}^H\mv{z}_{1,n}|^2 = 0$ for $\forall n$ by (\ref{con: norm0}), which means $\mv{H}\mv{Z}=\mv{0}$.
	Since $\mv{Z}$ is the orthogonal basis for the null space of $\mv{D}^*$, we can get
	\begin{align}
		\mv{S}^*\mv{Z}=\left(\mv{D}^*+(1+\lambda^*)\mv{H}\right)\mv{Z}=\mv{0}.
	\end{align}
	In addition, according to Lemma \ref{lemma 2} and ({\ref{con: S}}), we can get ${\rm{rank}}(\mv{S}^*)\geq {\rm{rank}}(\mv{D}^*)-{\rm{rank}}((1+\lambda^*)\mv{H})=l-1$. We define $\mv{\Omega}$ as the orthogonal basis for the null space of $\mv{S}^*$, which satisfies that
	\begin{align}
		{\rm{rank}}(\mv{\Omega})= N_t - {\rm{rank}}(\mv{S}^*)\leq N_t -l+1. \label{con: rankO}
	\end{align}
	We prove that ${\rm{rank}}(\mv{\Omega}) = N_t - l +1$ by following cases.
	\begin{itemize}
		\item Case 2.1: ${\rm{rank}}(\mv{\Omega})\geq N_t-l$. Since $\mv{Z}$ spans $N_t-l$ orthogonal dimensions of the null space of $\mv{S}^*$, we have ${\rm{rank}}(\mv{\Omega})\geq N_t-l$.
		\item Case 2.2: ${\rm{rank}}(\mv{\Omega})\neq N_t-l$. If ${\rm{rank}}(\mv{\Omega}) = N_t-l$, we can get $\mv{\Omega} = \mv{Z}$. Then, we have $\mv{W}^* = \sum_{n=1}^{N_t-l}a_n\mv{z}_{1,n}\mv{z}_{1,n}^H, a_n\geq 0$ for $\forall n$. However, $\mv{z}_{1,n}, \forall n$ lie in the null space of $\mv{H}$, which means there is no information transmitted to the user.
		\item Case 2.3: ${\rm{rank}}(\mv{\Omega})= N_t-l+1$. according to (\ref{con: rankO}), there exists only one single subspace spanned by $\mv{r}$ of unit norm, which lies in the null space of $\mv{S}^*$ and is orthogonal to the span of $\mv{Z}$. Thus, we have
		\begin{align}
			\mv{\Omega}=[\mv{Z}, \mv{r}],
		\end{align}
		where ${\rm{rank}}(\mv{\Omega})= N_t-l+1$. According to (\ref{con:proof1}) and (\ref{con: KKTs}), the optimal solution $\mv{W}^*$ can be expressed as
		\begin{align}
			\mv{W}^* = \sum_{n=1}^{N_t-l}a_n\mv{z}_{1,n}\mv{z}_{1,n}^H + b{\mv{r}}{\mv{r}}^H
		\end{align}
		where $a_n\geq0$ for $\forall n$, $b>0$, and $\mv{r}$ satisfies $\mv{r}^H\mv{Z} = \mv{0}$.
	\end{itemize}
\end{itemize}

The proof of the expressions of optimal solutions $\mv{W}^*$ and $\mv{V}^*$ for the problem (P2.1R) is completed.

3) We consider that $({\mv{\overline{W}}}^*, \overline{\mv{V}}^*, \overline{t}^*)$ given in (\ref{W}), (\ref{V}) and (\ref{t}) with ${\rm{rank}}(\overline{\mv{W}}^*) = 1$ is also an optimal solution to (P2.1R). Take (\ref{W}), (\ref{V}) and (\ref{t}) into (P2.1R), we have
\begin{align}
	{\rm{tr}}\left({\mv{H}}{\mv{\overline{W}}}^*\right) &= {\rm{tr}}\left({\mv{H}}\left({\mv{W}}^*-\sum_{n=1}^{N_t-l}a_n\mv{z}_{1,n}\mv{z}_{1,n}^H \right)\right)\\ \nonumber
	&={\rm{tr}}\left({\mv{H}}{\mv{W}}^*\right),
\end{align}
\begin{align}
	{\rm{tr}}\left({\mv{H}}{{\mv{\overline{V}}}}^*\right)+\overline{t}^*\sigma^2 &= {\rm{tr}}\left({\mv{H}}\left({\mv{V}}^*+\sum_{n=1}^{N_t-l}a_n\mv{z}_{1,n}\mv{z}_{1,n}^H \right)\right)\\ \nonumber+{t}^*\sigma^2
	&={\rm{tr}}\left({\mv{H}}{\mv{V}}^*\right)+{t}^*\sigma^2 = 1,
\end{align}
\begin{align}
	{\rm{tr}}\left({\mv{A}_k}{{\mv{\overline{W}}}}^*\right)&\leq{\rm{tr}}\left({\mv{A}_k}{{\mv{W}}}^*\right)\leq\gamma\left({\rm{tr}}\left({\mv{A}_k}{{\mv{V}}}^*\right)+\frac{{t}^*\sigma_E^2r^2}{\beta_0}\right)\\ \nonumber
	&\leq\gamma\left({\rm{tr}}\left({\mv{A}_k}{\overline{\mv{V}}}^*\right)+\frac{\overline{t}^*\sigma_E^2r^2}{\beta_0}\right), \quad\forall k
\end{align}
\begin{align}
	{\rm{tr}}\left({\overline{\mv{W}}}^*\right)+{\rm{tr}}\left({\overline{\mv{V}}}^*\right)&= {\rm{tr}}\left({{\mv{W}}}^*\right)+{\rm{tr}}\left({{\mv{V}}}^*\right)\leq \overline{t}^*P,
\end{align}
\begin{align}
	{\rm{tr}}\left(\left({\overline{\mv{W}}}^*+{\overline{\mv{V}}}^*\right){\mv{\bar{Q}}}\right)&= {\rm{tr}}\left(\left({{\mv{W}}}^*+{{\mv{V}}}^*\right){\mv{\bar{Q}}}\right)\\ \nonumber
	&\geq\frac{\overline{t}^*\sigma_R^2}{2|\bar{\beta}|^2}\left(\frac{1}{\Gamma}-\frac{1}{\sigma_\theta^2}\right),
\end{align}
\begin{align}
	{\rm{tr}}\left({\overline{\mv{W}}}^*\right)\geq{\mv{0}},\quad {\rm{tr}}\left({\overline{\mv{V}}}^*\right)\geq{\mv{0}},\quad {\overline{t}^*}\geq{0}.
\end{align}
Therefore, $({\mv{\overline{W}}}^*, \overline{\mv{V}}^*, \overline{t}^*)$ can be an optimal solution for (P2.1R) with ${\rm{rank}}(\overline{\mv{W}}^*)=1$.

Proposition \ref{pro 2} is thus proved.

\end{document}